\documentclass{article}
\usepackage{spconf,amsmath,graphicx}

\usepackage{enumitem}
\setlist{nosep, leftmargin=14pt}

\usepackage{mwe} 
\usepackage{multirow} 
\usepackage{makecell} 
\usepackage{textcomp}
\usepackage{gensymb} 
\usepackage{stmaryrd} 

\usepackage{sidecap} 
\usepackage{caption}
\usepackage{subcaption}

\title{Segmentation of the cortical plate in fetal brain MRI\\with a topological loss}

\name{\begin{tabular}{c}Priscille de Dumast$^{1, 2}$, Hamza Kebiri$^{1, 2}$, Chirine Atat$^{1}$\\ Vincent Dunet$^{1}$, Mériam Koob$^{1}$, Meritxell Bach Cuadra$^{1, 2}$\end{tabular}}

\address{$^{1}$  Department of Diagnostic and Interventional Radiology,\\ Lausanne University Hospital and University of Lausanne, Lausanne, Switzerland \\
$^{2}$ CIBM Centre d'Imagerie BioMédicale, Lausanne, Switzerland }

\begin{document}
\ninept
\maketitle
\begin{abstract}
The fetal cortical plate undergoes drastic morphological changes throughout early \textit{in utero} development that can be observed using magnetic resonance (MR) imaging. An accurate MR image segmentation, and more importantly a topologically correct delineation of the cortical gray matter, is a key baseline to perform further quantitative analysis of brain development. In this paper, we propose for the first time the integration of a topological constraint, as an additional loss function, to enhance the morphological consistency of a deep learning-based segmentation of the fetal cortical plate.
We quantitatively evaluate our method on 18 fetal brain atlases ranging from 21 to 38 weeks of gestation, showing the significant benefits of our method through all gestational ages as compared to a baseline method. Furthermore, qualitative evaluation by three different experts on 130 randomly selected slices from 26 clinical MRIs evidences the out-performance of our method independently of the MR reconstruction quality.
\end{abstract}
\section{INTRODUCTION}
\label{sec:intro}
The early \textit{in utero} brain development involves complex intertwined processes, reflected in both physiological and structural changes \cite{Tierney2009BrainDA}. The developing cortical plate specifically undergoes drastic morphological transformations throughout gestation. Nearly all gyri are in place at birth, even though the complexification of their patterns carries on after birth \cite{lenroot_2006}. T2-weighted (T2w) magnetic resonance imaging (MRI) offers a good contrast between brain tissues, hence allowing to assess the brain growth and detect abnormalities \textit{in utero}. In the clinical context, fetal MRI is performed with fast, 2D orthogonal series in order to minimize the effect of unpredictable fetal motion but results in low out-of-plane spatial resolution and significant partial volume effect. In order to combine these multiple series, advanced imaging techniques based on super-resolution (SR) algorithms \cite{tourbier_srtv_2015, ebner_niftymic_2020} allow the reconstruction of 3D high-resolution motion-free isotropic volumes. Together with improved visualization, these SR volumes open up to more accurate quantitative analysis of the growing brain anatomy. Consequently, based on 3D reconstructed volumes, multiple studies explored semi-automated fetal brain tissue segmentation \cite{makropoulos_review_segmentation} and cortical folding patterns \textit{in-utero} \cite{clouchoux_quantitative_2012, Wright2014AutomaticQO}. However, cortical plate segmentation remains challenging as it undergoes significant changes due to the brain growth and maturation, respectively modifying the morphology and the image contrast \cite{makropoulos_review_segmentation}. In this respect, we aim at developing a fully automatic and topologically correct age-invariant segmentation method of the cortical plate.
The first topological based clustering method for the segmentation of the fetal cortex was introduced in \cite{caldairou_2d_2011, caldairou_3d_2011}. This model relied on geometrical constraints to integrate anatomical and topological priors. Regrettably, their topological correctness was not further evaluated and results were presented qualitatively only for 6 fetuses. More recently, deep learning methods have also focused on fetal brain MRI cortical gray matter segmentation. Using a neonatal segmentation framework as initialization, \cite{fetit2020a} proposes a multi-scale approach for the segmentation of the developing cortex, while \cite{dou2020deep} implements a two-stage segmentation framework with an attention refinement module. Nevertheless, while the segmentation accuracy of these recent deep learning methods is promising, none of these works assesses the topological correctness of their results. In fact, these works report high overlap metrics but illustrated results show lack of topological consistency with notably discontinuous/broken cortical ribbons.
To overcome this limitation and further improve deep learning architectures for cortical plate segmentation in fetal MRI, in this paper we propose for the first time the integration of a topological constraint in a deep image segmentation framework. 

\begin{figure*}[ht!]
    \centering
    \includegraphics[width=0.9\linewidth]{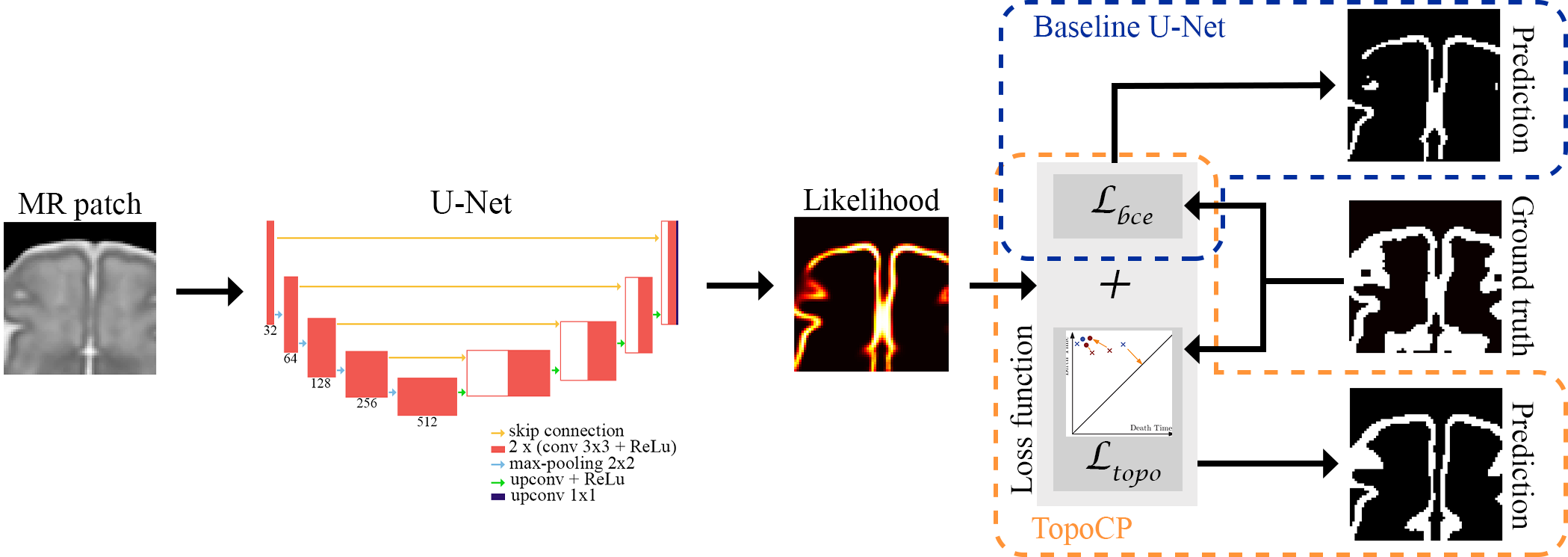}
    \caption{Figure inspired and adapted from \cite{TopologyPreserving2019}. Our method, TopoCP, integrates a topological loss based on persistent homology to a 2D U-Net segmentation of cortical plate fetal MRI.}
\label{fig:topoCP_framework}
\end{figure*}

\section{METHODOLOGY}
\label{sec:method}

\begin{table*}
\small
  \centering
\begin{tabular}{ c | c c c c c c } 
 Dataset  & \thead{Magnetic field \\strength (Tesla)}    & Vendor   & \thead{Image \\resolution (mm$^{3}$)} & \thead{Number \\of subjects} & \thead{Gestational\\age (weeks)} & \thead{Reconstruction\\method} \\ 
\hline 
TRAINING & 1.5T; 3T & General Electric & 0.5x0.5x0.5 & 15 & \thead{[22.6 - 33.4]\\ (28.7$\pm$3.5)} &  \thead{mialSRTK\\\cite{tourbier_srtv_2015, mialsrtk_mevislab}}\\ 
\thead{EVALUATION \\Quantitative} & 1.5T; 3T &  Siemens; Philips &  0.8x0.8x0.8 & 18 & 21-38 & \thead{Gholipour et. al, \\2017 \cite{gholipour_normative_2017}}\\ 

\thead{EVALUATION \\Qualitative} & 1.5T & Siemens & 1.12x1.12x1.12 & 26 & \thead{[18-34] \\ (27.3$\pm$4.1)} & \thead{mialSRTK\\\cite{tourbier_srtv_2015, mialsrtk_mevislab}}
\end{tabular}
\caption{Summary of the data used for training and quantitative and qualitative evaluation.}
\label{table:data_summary}
\end{table*}
\subsection{Topological loss}
\label{ssec:topological_loss}
Our approach, inspired from \cite{TopologyPreserving2019}, integrates a topological loss function into a baseline deep learning segmentation framework (as illustrated in Figure~\ref{fig:topoCP_framework}). This topology-preserving loss compares the predicted likelihood to the ground truth (GT) segmentation using the concept of persistent homology \cite{Edelsbrunner_topology_2000}. In a nutshell, homology structures are obtained by filtration to all possible threshold values of the prediction from the last layer of the network and reported in a persistence diagram. Both 0-dimensional and 1-dimensional Betti numbers \cite{gardner_betti_1984}, corresponding respectively to the number of connected components and the number of holes, are tracked. The persistence diagrams of the likelihood and the GT are matched, finding the best one-to-one structure correspondence, and the topological loss is computed as the distance between the matched pairs. 
\subsection{Network architecture}
\label{ssec:network_architecture}
The topological loss introduced above is indeed compatible with any deep neural network providing a pixel-wise prediction.
We chose as baseline the popular U-Net \cite{Ronneberger_Unet_2015} image segmentation method, as it recently proved its good ability to deal with fetal brain MRI tissue segmentation \cite{khalili_fetal_UNet_2019}. Thus, two different models are built and trained:
\begin{enumerate}
    \item \textit{Baseline 2D U-Net} is 
    trained using the binary cross-entropy function loss  \begin{math}\mathcal{L}_{bce}\end{math}; 
    \item \textit{TopoCP}, with the same architecture, is trained with the following loss combination: 
    \begin{equation}\mathcal{L}=\mathcal{L}_{bce}+\mathcal{L}_{topo}\end{equation} 
    where \begin{math}\mathcal{L}_{topo}\end{math} is the topological term presented in Section \ref{ssec:topological_loss}. 
\end{enumerate}
The U-Net architecture is composed of encoding and decoding paths.
The encoding path in our study is composed of 5 repetitions of the followings: two 3x3 convolutional layers, followed by a rectified linear unit (ReLu) activation function and a 2x2 max-pooling downsampling layer. Feature maps are hence doubled from 32 to 512. In the expanding path, 2x2 upsampled encoded features concatenated with the corresponding encoding path are 3x3 convolved  and passed through ReLu. The network prediction is computed with  final 1x1 convolution.

\subsection{Training strategy}
\label{ssec:training_strategy}
The publicly available dataset Fetal Tissue Annotation and Segmentation Dataset (FeTA) is used in the training phase \cite{FeTA2020_dataset, payette2020comparison}. Discarding pathological and non-annotated brains, our training dataset results in 15 healthy fetal brains (see details summarized in Table \ref{table:data_summary}).

Both networks are fed with 64x64 patches, containing cortical gray matter. Data augmentation is performed by randomly flipping and rotating patches (by $n\times90\degree$, $n\in\llbracket0;3\rrbracket$).
As in \cite{TopologyPreserving2019}, to overcome the high computational cost of persistent homology, we adopted the following optimization strategy: 1) our baseline model was trained over 23 epochs with a learning rate decay scheduled at epochs 11, 16, 17, 22 and a decay factor of 0.5, initialized at $0.0001$ ; 2) from the pretrained model in the first step, both networks were fine-tuned over 35 epochs, with learning rate decay scheduled at 14, 23 for Baseline U-Net and none for TopoCP. A cross-validation approach was used to determine the epochs for learning rate decay.
\begin{figure*}[ht!]
    \centering
    \includegraphics[width=0.8\textwidth]{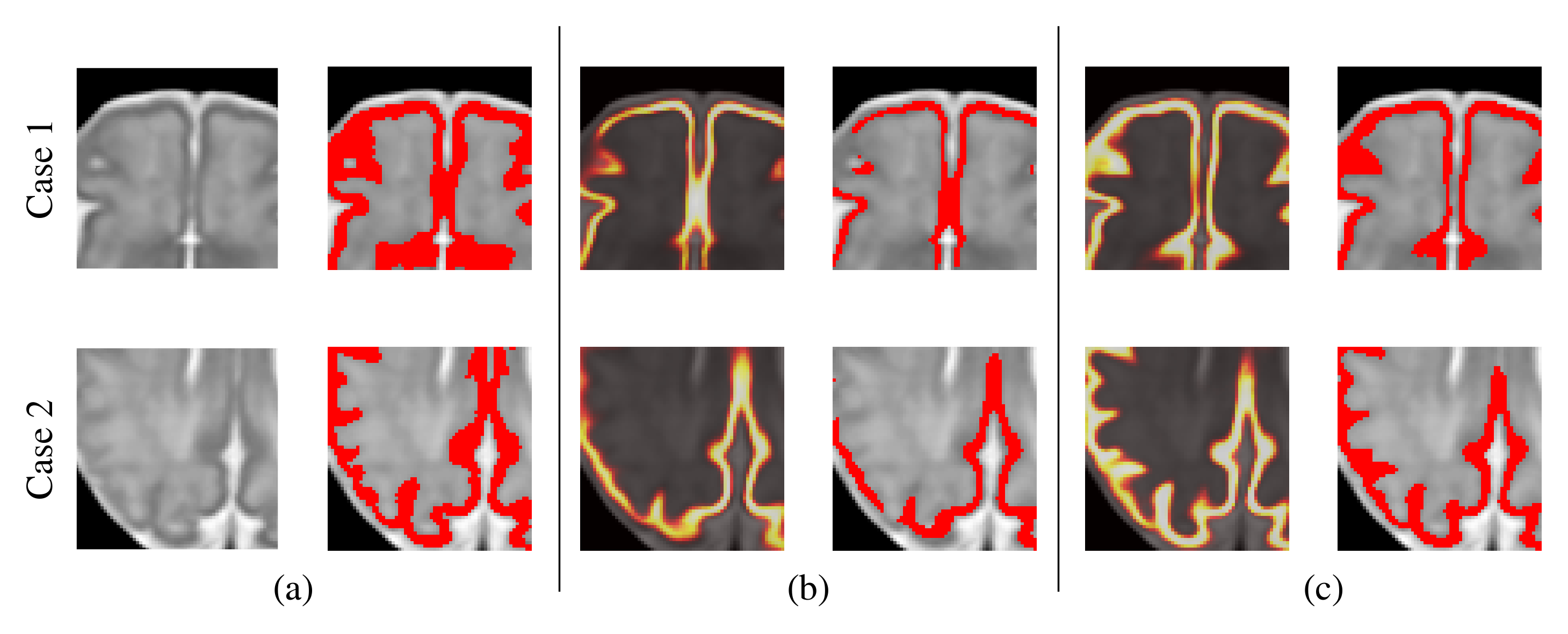}
    \caption{ Segmentation results on 35 GA atlas. (a) T2w (left) and GT segmentation overlaid (right). (b) Baseline U-Net and (c) TopoCP: predicted likelihood (left) and estimated segmentation (right). Likelihood probabilities: 0 \protect\includegraphics[height=0.6\baselineskip]{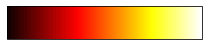} 1. Case 1 illustrates a net improvement in the segmentation of the midsagittal area and frontal cortical foldings. Case 2 shows a more accurate detection of the deep sulci with TopoCP.}
\label{fig:gholipour_visual_results_gt_lh_sg}
\end{figure*}
\begin{table*}
\centering
\small
\begin{tabular}{ c || c c | c c | c c | c c } 
     Method & \multicolumn{2}{c}{DSC$\uparrow$} & \multicolumn{2}{c}{ASD (mm)$\downarrow$} & \multicolumn{2}{c}{HD95 (mm)$\downarrow$} & \multicolumn{2}{c}{Betti Error$\downarrow$} \\
    \hline
    Baseline U-Net  & 0.54$\pm$0.16 & \multirow{2}{*}{\textit{$<$2e-16}} & 2.51$\pm$2.34 & \multirow{2}{*}{\textit{$<$2e-16}} & 7.32$\pm$6.57 & \multirow{2}{*}{\textit{$<$2e-16}} & 0.74$\pm$1.3 & \multirow{2}{*}{\textit{0.023}} \\ 
    TopoCP & \textbf{0.70$\pm$0.14} & & \textbf{1.41$\pm$1.64} & &  \textbf{5.31$\pm$5.20} & & \textbf{0.68$\pm$1.27}
\end{tabular}
\caption{Segmentation performances (mean$\pm$standard deviation) for both methods. Boldface text: best values for each metrics. Italic text: \textit{p}-values of paired Wilcoxon rank sum test adjusted with Bonferroni multiple comparisons correction, between both methods for each metric.}
\label{table:metrics_p_values_atlas_2}
\end{table*} 

\section{EVALUATION}
\label{ssec:evaluation}

\subsection{Quantitative Evaluation}
\label{ssec:quantitative_evaluation}

\textbf{Data.} In the training dataset (FeTA), label maps were sparse (annotations were performed on every 2$^{nd}$ to 3$^{rd}$ slice) and their interpolation resulted in \textit{noisy} labels showing topological inconsistencies. Therefore, we rather evaluate our method in an independent pure testing dataset, presenting topologically accurate segmentations. The normative spatiotemporal MRI atlas of the fetal brain \cite{gholipour_normative_2017} provides 3D high-quality isotropic smooth volumes along with tissue label maps for all gestational age between 21 and 38 weeks (see Table \ref{table:data_summary} for details).

\textbf{Analysis.} Performance of the methods is evaluated on patches extracted from the final prediction, using different types of metrics: 
1) the overlap between the GT and the predicted segmentation is quantified with the Dice similarity coefficient (DSC) \cite{dice_1945}; 
2) two boundary-distance-based metrics are measured to evaluate the contours: the 95th percentile of the Hausdorff distance (HD95) \cite{hausdorff_dist_1993} and the average symmetric surface distance (ASD) \cite{heimann_segmentation_evaluation_2009};
3) finally, the topological correctness is quantified with the Betti Error. Similarly to \cite{TopologyPreserving2019}, we define the latter as the average absolute difference of the 0-dimensional number of Betti (number of connected components).
To assess the significance of the observed differences between the two methods, we perform a paired Wilcoxon rank sum test for each metrics. \textit{p}-values were adjusted for multiple comparisons using Bonferroni correction and statistical significance level was set to 0.05.
\vspace{-0.1cm}
\subsection{Qualitative Evaluation}
\label{ssec:qualitative_evaluation}
\textbf{Data.} In order to better represent the diversity of the cortical variability and to prove the generalization of our approach to SR reconstructions of clinical acquisitions, we introduce a second pure testing set of T2w SR images of 26 healthy fetuses. Two subsets were created, from a consensus of three experts evaluation, based on the quality of the reconstructed 3D volumes: 1) excellent (N$=$16) and 2) acceptable (N$=$10) - with remaining motion artifacts or partial volume effects. Prior to the segmentation inference, clinical images were resampled to match the resolution of the training data, in order to present a similar field of view.

\textbf{Analysis.} Three raters (two radiologists and one engineer) performed a qualitative analysis of the baseline and TopoCP segmentations. Randomly-ordered segmentation of axial slices were presented. The experts were asked to indicate if they preferred either the segmentation A or B or if they were of equivalent quality. The inter-rater reliability was assessed with their percentage agreement before considering a consensus evaluation resulting from the majority voting of the experts' evaluations.

\section{RESULTS}\label{sec:results}
\begin{figure}[ht!]
    \centering
    \includegraphics[width=0.85\linewidth]{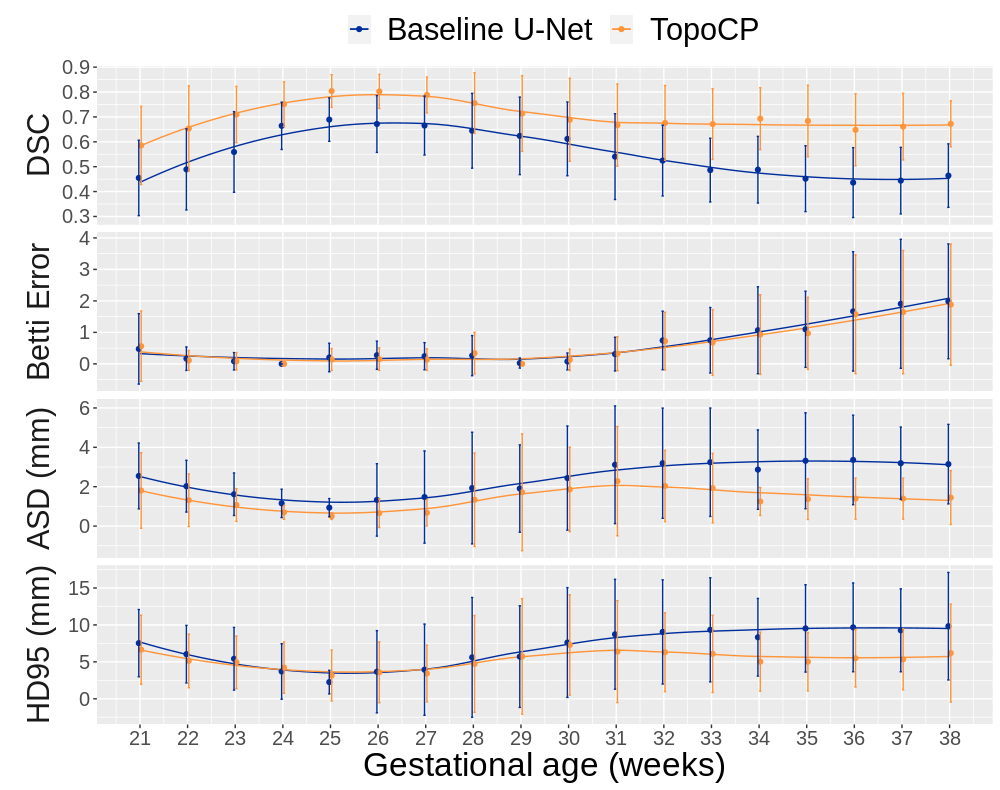}
\caption{Evolution of the performance metrics for atlas images from 21 to 38 weeks of gestation.}
\label{fig:gholipour_qualitative_results_fct_GA}
\end{figure}

Figure \ref{fig:gholipour_visual_results_gt_lh_sg} shows the GT of two representative patches with their predicted likelihood and segmentation overlaid on the T2w SR image. These results illustrate the benefits of TopoCP on the estimated probability maps, detecting more subtle variation of the cortex. The improved likelihood echoes with a better segmentation.
A summary of the performance metrics on the fetal brain atlas is presented in Table \ref{table:metrics_p_values_atlas_2}. TopoCP outperformed the Baseline U-Net in all evaluation metrics (highlighted in bold in the table). Corrected \textit{p}-values between both methods are shown in italic, indicating that our method significantly improves the segmentation.
Metrics are shown as a function of the subjects' gestational age in Figure \ref{fig:gholipour_qualitative_results_fct_GA}. While the overlap metric constantly improves throughout gestation, error distance and topological metrics improve mainly from the third trimester. We hypothesize this is due to the increased folding patterns of the cortical surface in the last trimester and thus where the topological constraint is more valuable.

\begin{figure} 
    \centering
    \includegraphics[width=0.9\linewidth]{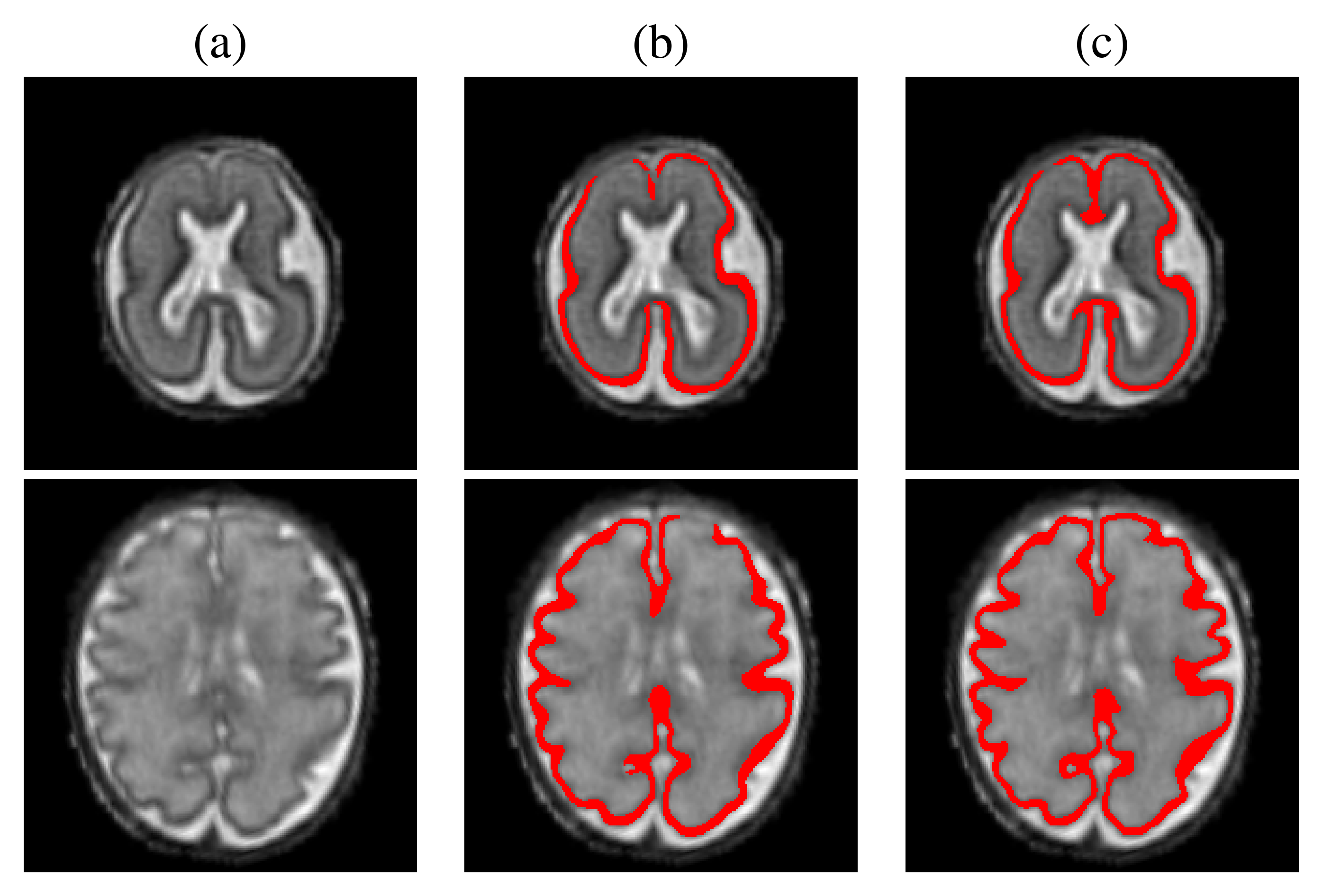}
    \caption{Segmentation results on 23 (top) and 32 (bottom) gestational weeks fetuses: (a) T2w SR image; (b) Baseline U-Net (c) and TopoCP.}
    \label{fig:chuv_qualitative_resuls}
\end{figure}

The inter-rater reliability showed a good agreement of 74\%. Table \ref{table:qualitative_eval_chuv} shows the consensus of the experts' blind evaluation of the cortical plate segmentation on SR volumes based on T2w clinical acquisitions. For both excellent and acceptable sets, TopoCP was selected as giving the best segmentation, showing the robustness of our method to the SR quality.
Figure \ref{fig:chuv_qualitative_resuls} illustrates a representative slice segmented with both methods.

\section{CONCLUSION}
\label{sec:conclusion}
The integration of a topological constraint improves the topological correctness of the segmentation of the fetal cortical plate on MRI. 
We evidenced the benefits of this additional topological loss in a 2D approach that can be easily extended to 3D if enough training subjects are available. We emphasize the genericity of this loss, which can be applied to any segmentation network providing a pixel-wise prediction. We believe that pairing the topological loss to state-of-the-art methods would considerably improve the resulting segmentation, even in a multi-class task.
\begin{table}
\centering
\small
\begin{tabular}{ c | c c c  } 
    \thead{SR quality\\(\# of slices)}  & Baseline U-Net    & TopoCP   & Equal \\ 
    \hline
    Acceptable (50)  & 12\%            & \textbf{84\%}   & 4\% \\ 
    Excellent (80)   & 17\%            & \textbf{79\%}   & 4\% \\
    All (130)        & 15\%            & \textbf{81\%}   & 4\%  
\end{tabular}
\caption{Experts' qualitative evaluation results in the comparison of Baseline U-Net and TopoCP automatic segmentations.}
\label{table:qualitative_eval_chuv}
\end{table}

\bibliographystyle{IEEEbib}
\bibliography{refs}

\clearpage

\section{ACKNOWLEDGMENTS}
\label{sec:acknowledgments}

This work has been supported by the Centre d’Imagerie BioMédicale (CIBM) of the University of Lausanne    (UNIL), the Swiss Federal Institute of Technology Lausanne (EPFL), the University of Geneva (UniGe), the Lausanne University Hospital (CHUV), the Hôpitaux Universitaires de Genève (HUG), and the Leenaards and Jeantet Foundations. This work is supported by the Swiss National Science Foundation (FNS projects 205321\_182602).
We thank Thomas Yu for the proof reading of the paper. 
The authors have no relevant financial or non-financial interests to disclose.

\section{Compliance with Ethical Standards}
\label{sec:ethics}

Publicly available datasets were used in this study: the normative spatiotemporal MRI atlas of the fetal brain\footnote{http://crl.med.harvard.edu/research/fetal\_brain\_atlas/} and Fetal Tissue Annotation and Segmentation Dataset\footnote{DOI: 10.18112/openneuro.ds003105.v1.0.1}.
Anonymized clinical data used in the evaluation of our method were part of a larger research protocol at our institution approved by the local ethics committee.

\clearpage

\vfill
\pagebreak

\end{document}